\documentclass{mem}
\usepackage{graphicx}
\usepackage{times}
\usepackage{balance}
\usepackage{natbib}
\usepackage{txfonts}
\usepackage[a4paper]{hyperref}

\begin{document}

\title{Merger and non-merger galaxy clusters in cosmological AMR simulations.}

 \subtitle{}
\author{
Vazza F.\inst{1,2}}
\offprints{\email{f.vazza@jacobs-university.de}}
\institute{
Jacobs University Bremen, Campus Ring 1, 28759, Bremen, Germany  \and INAF/Istituto di Radioastronomia, via Gobetti 101, I-40129 Bologna}



\abstract{{\it Aims}: We discuss the dependence of shocks, cosmic rays acceleration and turbulence on the dynamical state of the host clusters. 
{\it Method:} We perform cosmological simulations with the grid code ENZO 1.5, with a mesh refinement scheme tailored to follow at high resolution shocks and turbulence developed in the clusters volume.
{\it Results:} Sizable differences are found when some important properties, connected to non-thermal activity in clusters, are compared for post-merger, merging and relaxing systems.


}

\titlerunning{Merger and non-merger clusters}

\authorrunning{Vazza F.}

\maketitle

\begin{figure}
\begin{center}
\includegraphics[width=0.495\textwidth,height=0.495\textwidth]{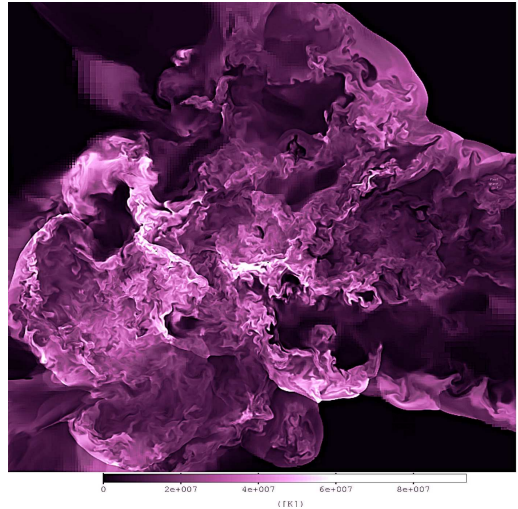}
\caption{Map of gas temperature for a major merger at $z \approx 0.6$. The side of the image is 8.8 Mpc/h, the depth along the line of sight is 25kpc/h.}
\label{fig:turbo_map}
\end{center}
\end{figure}

\section{Introduction}

\label{sec:intro}


Numerical simulations presently provide a unique way to study the generation and
evolution of shock waves, turbulence and chaotic motions following the evolution
of large scale cosmic structure over a wide range of scales, and in a fully-time
dependent way (\citealt{do08} and references therein). Simulations can also be a very powerful approach to tackle the problem of explaining the occurrence and morphologies of non-thermal emissions in clusters, observed in many systems (e.g. \citealt{fe08} for a review).
We recently employed a tailored Adaptive Mesh Refinement (AMR) scheme in the ENZO cosmological code (e.g. \citealt{osh04}; \citealt{no07}) in order to study these phenomena
with unprecedented high dynamic range (\citealt{va10a}; \citealt{va10b}). 
Here we summarize some of the most important findings of our work, reporting dependences between shocks, CR acceleration and turbulence and the dynamical state of the simulated host clusters.

\section{ ENZO Simulations.}
\label{sec:enzo}

The computations presented in this work were performed using the 
ENZO 1.5 code developed by the Laboratory for Computational
 Astrophysics at the University of California in San Diego 
(http://lca.ucsd.edu), see also \citet{osh04} and \citet{no07}.

We performed non-radiative $\Lambda$CDM simulations sampling a total cosmological volume with the size of $L_{\rm box} \approx 440$ Mpc/h, in which  we re-simulated the evolution of the 20 most massive galaxy clusters, with a DM  mass resolution of 
$m_{\mathrm DM}=6.76 \cdot 10^{8} M_{\sun}$ and a peak spatial resolution of 
$\Delta x \approx 25$ kpc/h inside $<3 R_{\rm vir}$ for each cluster (\citealt{va10a}).
Our Adaptive Mesh Refinement (AMR) strategy was tailored to keep the maximum available resolution also at 
large distances from the cluster centers, tracking the propagation of
strong discontinuities in the velocity field associated with shocks or turbulent motions (see \citealt{va09} for a detailed discussion). At $z=0$ each simulated clusters is sampled with a high number of cells, $N_{grid} \sim 500^{3}-600^{3}$, allowing a study of the intra cluster medium (ICM) across 2-3 orders of magnitude in spatial scales (see Fig. \ref{fig:turbo_map} for a representative example).  

Our clusters have total masses in the range $6 \cdot 10^{14} \leq M/M_{\sun}  \leq 3 \cdot 10^{15}$; they were
 divided according to their dynamical
state, following in detail their matter accretion history for $z<1.0$ (see \citealt{va10a} for details). 
According to this definition, our sample contains
10 post-merger systems (i.e. clusters with a merger with a mass ratio larger than $1/3$ for $z \leq 1$), 
6 merging clusters and 4 relaxed clusters at $z=0$.

\begin{figure}
\includegraphics[width=0.45\textwidth,height=0.41\textwidth]{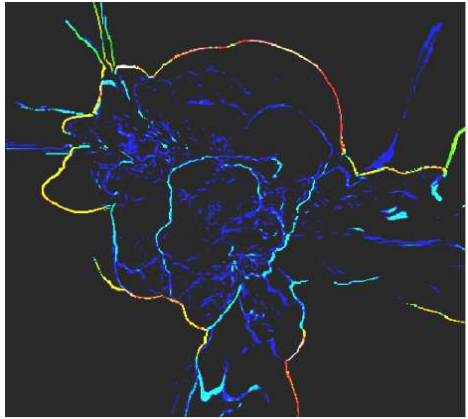}
\caption{2-dimensional slice showing the Mach number of shocks for cluster E1 of our sample. The side of the slice is 13 Mpc/h and the depth along the line of sight is 25 kpc/h.}
\label{fig:shock_map}
\end{figure}

\begin{figure}
\includegraphics[width=0.43\textwidth,height=0.4\textwidth]{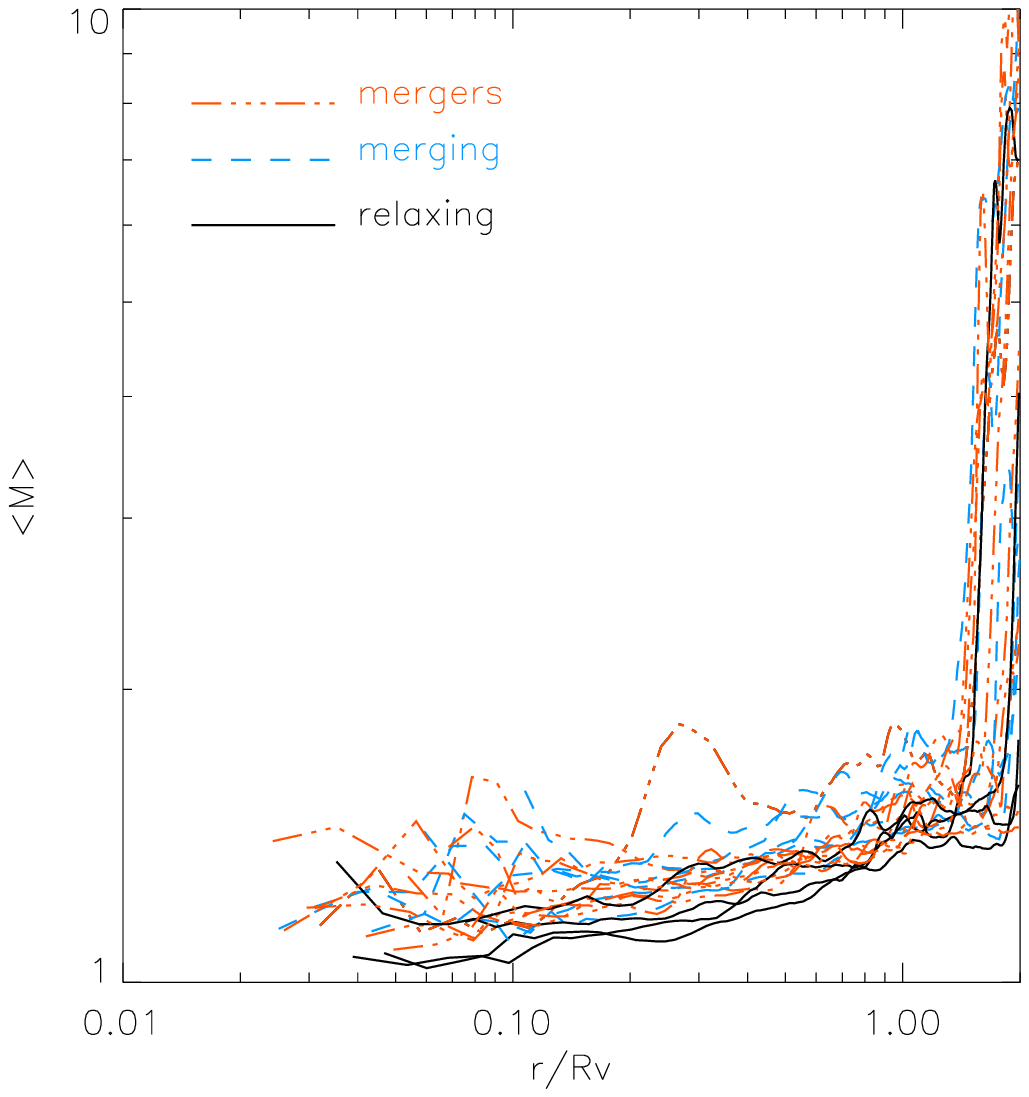}
\includegraphics[width=0.43\textwidth,height=0.4\textwidth]{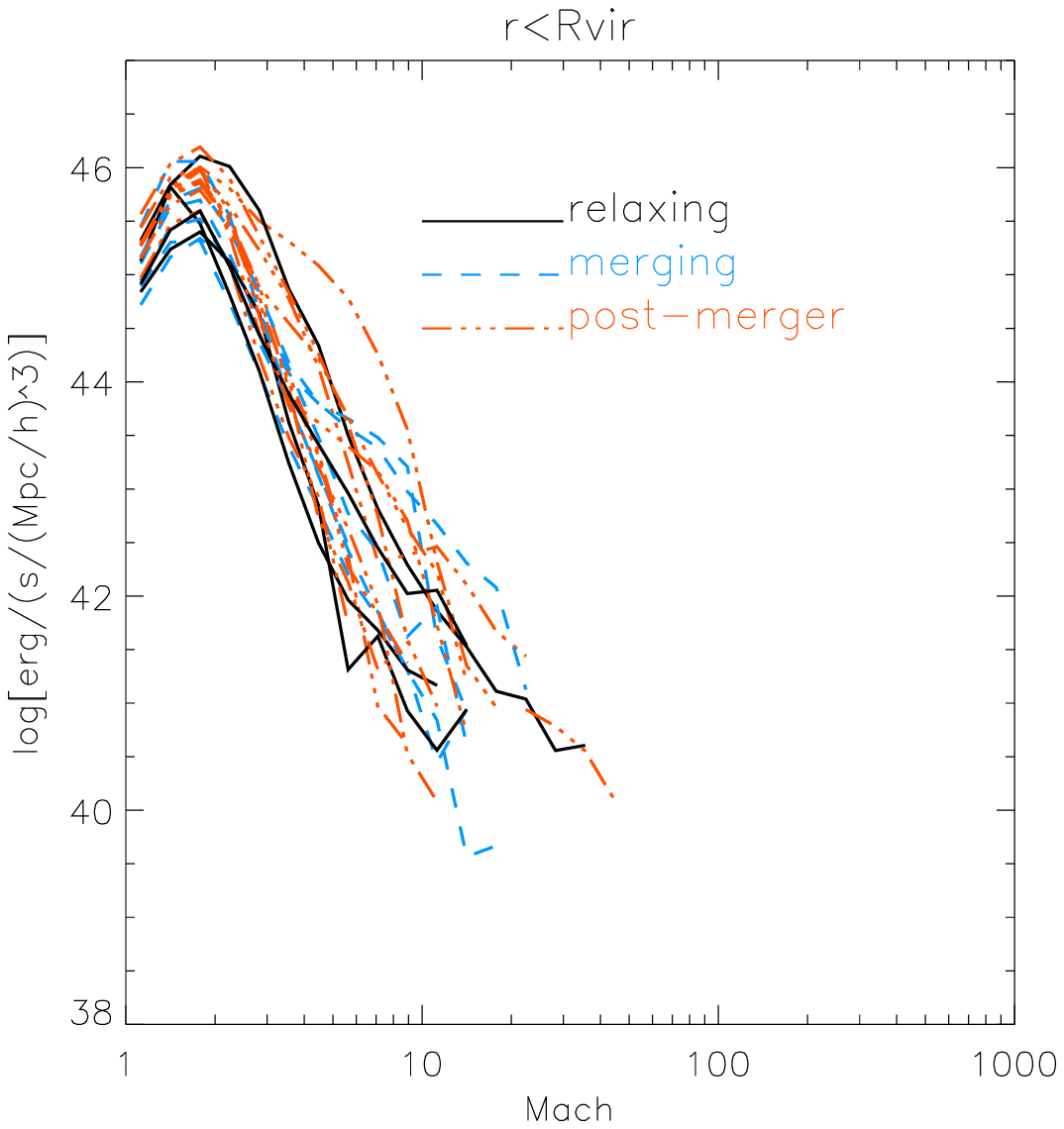}
\caption{Top: radial profile of average Mach number for the shocks in our sample. Bottom: distribution of thermal energy flux at shocks in the cluster samples.}
\label{fig:shocks_stat}
\end{figure}

\section{RESULTS}

\subsection{Shocks and Cosmic Rays}
\label{subsec:shocks}

We identified shocks in the ICM with the same procedure presented in \citet{vbg09}, 
based on the analysis of velocity jumps across close cells (see Figure \ref{fig:shock_map} for a map of the reconstructed Mach numbers).
All clusters show similar shocks frequencies and energy distribution inside $R_{\rm vir}$: the profiles of average Mach numbers are very flat and characterized by very weak
shocks, $M<2$ (top panel of Fig.\ref{fig:shocks_stat}). {\it Sizable and stronger shocks ($M \sim 2.5-3.5$) are detected only in one post-merger systems and in one merging system}. 
The volume distribution of shocks is very steep ($\alpha \leq -4$  with $\alpha = d\log N(M)/d\log M$), while the
distribution of flux of thermal energy through shock surfaces shows a well defined peak of thermalization at $M \sim 2$ (see bottom
panel of Fig.\ref{fig:shocks_stat}). In general our distributions are slightly steeper than what was previously reported in the literature (e.g. \citealt{ry03}; \citealt{pf06}; \citealt{sk08}), which can be explained as an effect of different resolution, re-ionization models and shocks detection schemes. Only for the few $M>2$ cases reported above the flux distribution is flatter, with $\alpha \sim -2$. 
We investigated the  injection
of relativistic protons (CR) at shocks applying the efficiency function introduced by \citet{kj07} in the framework of the Diffusive Shock Acceleration model. 
We found that the ratio between the CR energy flux and the thermal flux is $< 0.05$ inside $0.2R_{\rm vir}$ and $< 0.1$ inside $R_{\rm vir}$. {\it Only in one strong post-merger system 
we measure a CR injection of $\sim 0.2-0.3$ of the thermal energy flux.} The estimated amount of CR energy inside clusters
implied by our results is well below $\sim 0.1$, which is presently allowed by
observational upper limits coming from the non detection of gamma radiation from secondaries (e.g. \citealt{al10}) and from the non detection of diffuse radio emission in relaxed clusters (\citealt{br07}).

\begin{figure}
\includegraphics[width=0.45\textwidth,height=0.4\textwidth]{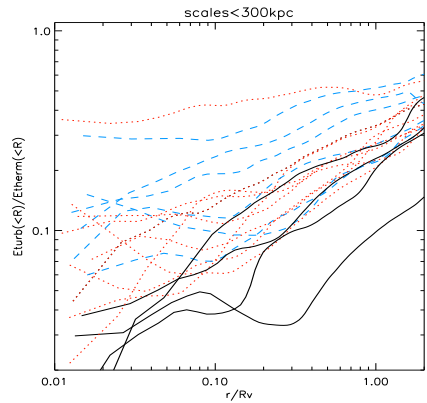}
\caption{Radial profile of the ratio between turbulent energy inside a given radius, and the thermal energy. The colors are as in Fig.\ref{fig:shocks_stat}.}
\label{fig:turbo_prof}
\end{figure}

\begin{figure*}
\begin{center}
\includegraphics[width=0.95\textwidth,height=0.3\textwidth]{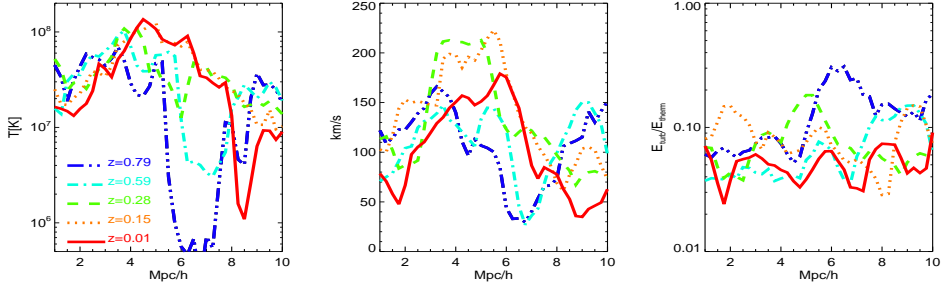}
\caption{Evolution of the profiles of gas temperature, turbulent gas velocity and $E_{\rm turb}/E_{\rm therm}$ along the axis of merger of two colliding clusters.}
\label{fig:turbo_evol}
\end{center}
\end{figure*}

\subsection{Turbulence}
\label{subsec:turbo}

\begin{figure}
\includegraphics[width=0.49\textwidth,height=0.4\textwidth]{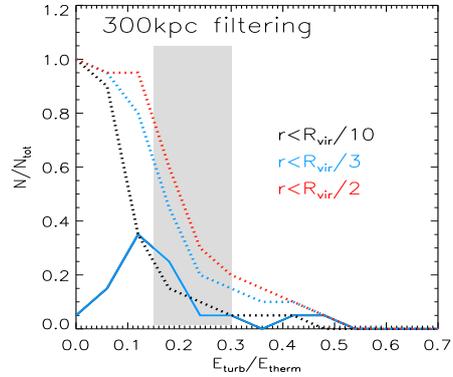}
\caption{Cumulative distribution function for $E_{\rm turb}/E_{\rm therm}$ inside $R_{\rm vir}/2$, $R_{\rm vir}/3$  and $R_{\rm vir}/10$ for
the simulated clusters assuming $l<300 kpc$ for turbulence. The solid line shows the differential distributions for $R_{\rm vir}/3$. The grey band shows the turbulence required in the turbulent re-acceleration
scenario (e.g. \citealt{bl07}).} 
\label{fig:turbo_halos}
\end{figure}

\begin{figure}
\includegraphics[width=0.475\textwidth,height=0.39\textwidth]{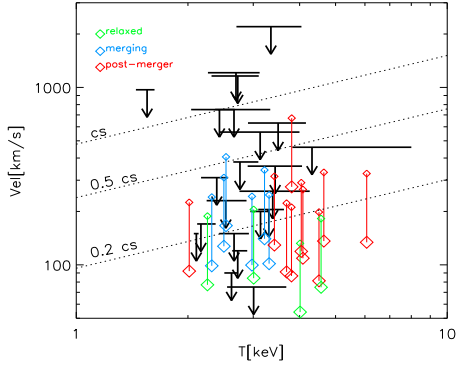}
\caption{Relation between temperature and turbulent velocity dispersion
(small squares) for our clusters (squares in colors) and
for the XMM-Newton observations of \citet{sa10}. The thick squares are for  turbulent field below $<30kpc$.}
\label{fig:turbo_limits}
\end{figure}

The turbulence in the ICM is expected to be sustained by the hierarchical process of matter accretion in evolving galaxy clusters (e.g. \citealt{bn99}; \citealt{do05}; \citealt{in08}).  
In \citet{va10b} we discussed the application of 2 different filtering techniques to disentangle the 
local mean velocity field of the gas (assuming it is laminar above a certain scale) 
and the turbulent velocity field. Despite small differences from case to case, the two methods provided consistent estimates of the turbulent energies of our clusters (see Sec.3.1 of \citealt{va10b}).
In Fig.\ref{fig:turbo_prof} we show the radial profile of the turbulent to
thermal energy ratio (for the filtering at the scale of $\approx 300$~kpc) for
all clusters in our sample. As expected {\it turbulent motions are strong
 in merging and in post-merger systems} ($E_{\rm turb}/E_{\rm therm} \sim 0.1-0.3$ for $r<0.2 R_{\rm vir}$), while they
are rather weak within the cores of relaxed systems ($E_{\rm turb}/E_{\rm therm} < 0.05$). However, almost all clusters host a sizable amount of turbulence ($\sim 20-40 E_{\rm therm}$) within the total
virial volume. 
An interesting finding of our simulations is that, on average, {\it the turbulent energy present in merging systems is slightly larger than the turbulent energy of post-merger systems}, when normalized to the thermal energy of the host cluster. This happens because the thermal 
energy inside the two clusters is initially lowered as an effect of the accretion of cold gas from filaments, while the turbulent energy almost stadely increases (see panels of Fig.\ref{fig:turbo_evol} and also the discussion in Sec.3.2 of \citealt{va10b}).
Our sample is large enough to allow the statistical study of the 
occurrence of turbulence in cluster cores, and to compare with some basic
expectation from the theoretical "turbulent re-acceleration" (e.g. \citealt{gb01}), as for instance with the estimated frequency of "turbulent" clusters. We investigated this issue in detail, by considering the typical volumes and turbulent
energies needed to produce realistic radio halo emission: 
at $z \sim 0$, about $1/3$ of our simulated clusters host a level of turbulent 
energy of $\sim 0.25 E_{\rm therm}$  inside a volume of $\sim R_{\rm vir}/3$, in line with existing observations (Fig.\ref{fig:turbo_halos}). 
Despite these rather large values of turbulence
in merging or post-merger clusters, the ICM at the smallest scale is only weakly turbulent, as follows from the general shape of the velocity power spectrum of our simulated ICM (\citealt{vgb10}; \citealt{xu09}). 
If turbulence in our clusters is computed for the typical spatial scales of real X-ray observations performed with XMM-Newton (e.g. \citealt{sa10}), $\sim 30$  kpc, the simulated turbulent velocities of the ICM within the core region of clusters are much smaller than the upper limits from observations (see 
Fig.\ref{fig:turbo_limits}).

\subsection{Azimuthal Scatter of Gas Profiles}
\label{subsec:scatter}

\begin{figure}
\includegraphics[width=0.45\textwidth,height=0.35\textwidth]{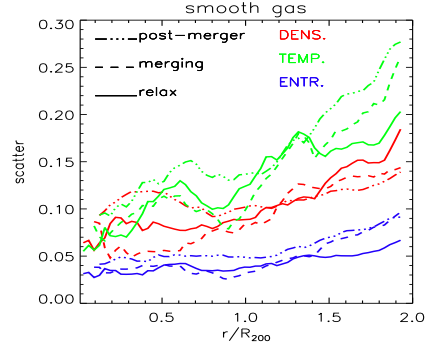}
\caption{Azimuthal scatter profile for our sample, after the "99 percent" filtering of the phsyical fields. The different colors refer to 
the different physical quantities, the linestyles refer to the dynamical classes of the sample.} 
\label{fig:scatter}
\end{figure}

We investigated the level of intrinsic azimuthal scatter from the center of
our clusters, focusing on some important thermodynamical quantities (e.g. density, temperature, entropy and X-ray luminosity).  This is interesting because recent long SUZAKU exposures measured sizable differences between the cluster profiles in different sectors from the center (e.g. \citealt{ba09}; \citealt{ge09}). The inspection of the radial properties of our simulated data offers a way to assess the need of non-gravitational physics in simulations, such as AGN, Cosmic Rays, magnetic field, in order to reconcile with observations. 
In Vazza et al. (2010~c) we tackled this issue by analyzing 
ENZO and GADGET2 runs of comparable resolution (\citealt{do05}). We processed all clusters with the filtering technique introduced in \citet{ro06}, designed to remove the densest clumpy gas component from each cluster dataset, and computed the cluster profiles in sectors of different angular size.
In Fig.\ref{fig:scatter} we report the azimuthal scatter ($\Delta \alpha=\pi/8$) for the different quantities in the ENZO clusters. {\it Post-merger clusters present the largest degree of azimuthal scatter at all radii while the relaxed ones have the smallest scatter}. The scatter increases of a factor $\sim 2$ going from $R_{\rm vir}$ to $2 \cdot R_{\rm vir}$; the scatter in temperature is always larger than the scatter of the other quantities. Once that the contribution of clumpiness is removed from the datasets, post-merger clusters are the one more affected by an asymmetric distribution of the thermal gas, due to the chaotic pattern induced along mergers and filamentary accretions. The
amount of azimuthal scatter is found to be in line with existing observations (see \citealt{va10c} for a detailed discussion), even if the presence of a un-removed clumpy component in the observation should be assumed to fit the case of cluster PKS0745-191 (\citealt{ge09}).  These results suggest that non-gravitational processes may be not needed in simulations to model the thermodynamics of the gas in the outer cluster regions (see also \citealt{bu10}; \citealt{na11}).

\section{Discussion and Conclusion.}
\label{sec:conclusions}


In summary, our AMR simulations with ENZO show that, for all (or most) of clusters:

\begin{itemize}
\item the volume and energy flux distribution of shocks in the ICM are steep and most of shocks have $M<2$;

\item the radial distribution of Mach numbers  is very flat inside $R_{\rm vir}$;

\item the average CR injection efficiency injection is small, $\epsilon_{\rm CR}/\epsilon_{\rm therm} < 0.05$;

\item the kinetic energy of turbulent motions is $\sim 0.2-0.4$ of the total thermal energy inside $R_{\rm vir}$;


\item the turbulent motions of the ICM at
scales $<30$  kpc are subsonic $v_{\rm turb} < 0.1 c_{\rm c}$;

\item the azimuthal scatter of the clusters profiles increases by $\sim 2$ from $R_{\rm vir}$ to $2 R_{\rm vir}$.
\end{itemize}

We find, however, the following significant dependences on the dynamical state of clusters:

\begin{itemize}
\item strong $M \sim 2.5-3.5$ shocks are found only inside the central region of clusters in $\sim 1/10$ of cases (never in relaxed clusters); 

\item in these rare systems, the injection efficiency of CR can be quite large, $\epsilon_{\rm CR} \sim 0.1-0.2 \epsilon_{\rm therm}$;

\item in post-merger system, the turbulent energy inside $R_{\rm vir}/3$ is $E_{\rm turb} \sim 0.25 E_{\rm therm}$, while it is  $\sim 0.1  E_{\rm therm}$ in post-merger systems and  $\sim 0.05 E_{\rm therm}$ in relaxed ones;

\item post-merger systems present a larger volume filling factor of turbulent motions compared to relaxed systems, and host enough turbulent energy to explain radio-halos within the re-acceleration scenario (see Brunetti et al. 2007);


\item the azimuthal scatter in merging and post-merger clusters is larger (by a $20-50$ percent) compared to the relaxed ones.

\end{itemize}

\section{Acknowledgments}
 
I gratefully acknowledge my advisors and collaborators, who made all these works possible: G. Brunetti, C. Gheller, M. Br\"{u}ggen, R. Brunino, M. Roncarelli, S. Ettori and K. Dolag. I also thank M. Br\"{u}ggen for carefully reading the paper.

I acknowledge partial 
support through grant ASI-INAF I/088/06/0 and PRIN INAF 2007/2008, and grant FOR1254 from Deutschen Forschungsgemeinschaft. The simulations were produced thanks to the CINECA-INAF 2008-2010 agreement.


\bibliographystyle{aa.bst}
\bibliography{vazza_1.bib}

\end{document}